\documentclass[aip,pop,reprint,amsmath,amssymb,floatfix,nofootinbib]{revtex4-2}

\usepackage{graphicx}
\usepackage{xcolor}
\usepackage[colorlinks=true,linkcolor=blue,urlcolor=blue,citecolor=blue]{hyperref}
\usepackage[nameinlink,capitalise]{cleveref}
\crefname{figure}{Fig.}{Figs.}

\begin{document}

\title{Bayesian optimization of double-pulse temporal shaping for enhanced
target-normal-sheath proton acceleration under fixed laser energy}

  \author{Cheng-Qi Zhang}
	\affiliation{
	Key Laboratory of Beam Technology of the Ministry of Education, and School of Physics and Astronomy, Beijing Normal University, Beijing 100875, China}

	\author{Yang He }
	\affiliation{Xinjiang Key Laboratory of Solid State Physics and Devices,
	School of Physics Science and Technology, Xinjiang University, Urumqi 830017, China}
	\author{Mamat Ali Bake }
	\affiliation{Xinjiang Key Laboratory of Solid State Physics and Devices,
	School of Physics Science and Technology, Xinjiang University, Urumqi 830017, China}
	\author{Bai-Song Xie}\thanks{Corresponding author. Email: bsxie@bnu.edu.cn}
	\affiliation{
	Key Laboratory of Beam Technology of the Ministry of Education, and School of Physics and Astronomy, Beijing Normal University, Beijing 100875, China}

\begin{abstract}
Splitting an ultrashort drive pulse into a weak leading pulse and a strong main pulse is known to raise the energy of protons accelerated by the
target-normal-sheath-acceleration (TNSA) mechanism, because the leading pulseforms a preplasma that increases the absorption of the main pulse. The
allocation of energy between the two pulses and their temporal separation are coupled control parameters, and under a fixed total energy they have not been
optimized jointly in a systematic way. We address this problem with two-dimensional particle-in-cell simulations driven by Bayesian optimization.
Treating the prepulse energy fraction $r$ and the interpulse delay $\Delta t$ as free parameters under a fixed total energy, a campaign of 32 simulations, of
which 16 are Sobol-initialized and 16 adaptively selected, locates an optimum at
$r\approx0.07$ and $\Delta t\approx234$~fs. The proton cutoff energy increases
from 7.7~MeV for the single pulse to 17.7~MeV at the optimum, a gain of about
130\%. The optimum is asymmetric with only about 7\% of the energy in
the leading pulse. At the
optimum the laser absorption rises from 4.84\% to 20.09\%, the bulk hot-electron
temperature from 1.20 to 1.94~MeV, and the time-integrated rear sheath field by a
factor of about 1.7. The optimum lies on a broad plateau in $\Delta t$, which
relaxes the timing tolerance required in an experiment. 
\end{abstract}

\maketitle

\section{Introduction}
Laser-driven proton sources are of interest for applications that include proton
radiography~\cite{borghesi2002macroscopic}, isochoric heating, fast ignition of
inertial fusion~\cite{roth2001fast}, and ion
therapy~\cite{linz2007will,bulanov2002feasibility,kroll2022tumour}. Their
development has followed the steady increase in available laser peak power at
petawatt-class facilities~\cite{danson2019petawatt}. Among the acceleration
mechanisms, target normal sheath acceleration (TNSA) is the most widely used
because of its robustness and relatively modest requirements on laser contrast
and target quality~\cite{snavely2000intense,wilks2001energetic,macchi2013ion,daido2012review,gibbon2005short}.
Other schemes such as radiation pressure
acceleration~\cite{esirkepov2004highly} and transparency-enhanced hybrid
acceleration~\cite{higginson2018near} can reach higher energies but place
stronger demands on the laser and the target. Cutoff energies above 85~MeV have
been reported for TNSA from thin foils~\cite{wagner2016maximum}.

In TNSA, the laser pulse heats electrons at the target front, a fraction of which
cross the foil and form a charge-separation sheath at the rear surface. The
resulting field ionizes and accelerates protons from surface
contaminants~\cite{hatchett2000electron,mora2003plasma,passoni2010target}. The
proton cutoff energy is set by the chain that connects laser
absorption, hot-electron temperature, and the strength and lifetime of the rear
sheath~\cite{fuchs2006laser,schreiber2006analytical,mora2005thin}.
The front-surface state at pulse peak arrival largely governs the subsequent energy coupling.  A controlled density gradient such as a preplasma at the
front surface increases absorption and raises the hot-electron temperature.
Measurements and simulations have shown that the
prepulse level, and hence the front-surface scale length, has an influence on the maximum proton energy~\cite{kaluza2004influence,mckenna2008effects,beg1997study}.
A practical way to impose such a gradient deliberately is to precede the main
pulse with a weaker drive pulse. Markey \textit{et al.} reported an
increase in the maximum proton energy and in the conversion efficiency to fast
protons when the drive energy was divided between two pulses separated in
time~\cite{markey2010spectral}. In that experiment, carried out on the VULCAN
laser\cite{hernandez2010vulcan}, the two sub-picosecond pulses were derived from the same system and
focused sequentially onto a single target with a controllable delay. A
pulse can be divided by a birefringent element or by a split-and-delay
interferometer~\cite{fraggelakis2019double}, and the addition of a half-wave
plate before a polarizing beam splitter sets the energy ratio of the two pulses
continuously while a delay stage sets their separation. Related multi-pulse
schemes have been studied numerically, including colliding pulses incident at
opposite angles~\cite{ferri2019enhanced} and double pulses with varied intensity
and duration~\cite{kumar2020optimization}.

These studies establish that preplasma-enhanced absorption can raise proton
energies. The open question is the joint optimization of the two pulses under a
fixed total drive energy. This constraint couples the two control parameters,
the energy split $r$ and the interpulse delay $\Delta t$. Energy diverted into
the prepulse is removed from the main pulse, weakening the drive of the rear
sheath. In exchange, the prepulse forms a front-surface preplasma whose density
gradient enhances absorption and thereby raises the hot-electron temperature, so
the preplasma in turn favors a higher cutoff energy. The delay sets how far this
preplasma expands before the main pulse arrives, and hence the scale length the
main pulse encounters. A delay that is too short leaves no beneficial gradient,
whereas one that is too long allows the front surface to over-expand and degrade
the coupling. The optimum thus reflects a balance in which the conditioning gain
from the prepulse must outweigh the acceleration energy it sacrifices. This
balance is not evident, which motivates a systematic search for the pair $(r,\Delta t)$ that maximizes the proton cutoff energy.

Mapping this trade-off by a dense parameter scan is expensive, because each
evaluation is a full kinetic simulation. Bayesian optimization (BO) is a
sample-efficient strategy for such expensive black-box
problems~\cite{jones1998efficient,shahriari2016taking,snoek2012practical}. It
builds a probabilistic surrogate of the objective and uses an acquisition
function to choose the next evaluation, balancing exploration of uncertain
regions against exploitation of promising
ones~\cite{rasmussen2006gaussian,srinivas2010gaussian}. The approach has been
adopted across laser-plasma physics, including laser wakefield
accelerators~\cite{shalloo2020automation,jalas2021bayesian} and laser-driven ion
acceleration~\cite{dolier2022multi,loughran2023automated}. Dolier \textit{et al.} optimized laser and target parameters (laser energy, pulse duration, target thickness, and front-surface scale length) to achieve maximized laser-driven ion acceleration in particle-in-cell simulations, and found a nontrivial interior optimum specifically for the front-surface plasma density scale length~\cite{dolier2022multi}.

In this work, we apply Bayesian optimization to two-dimensional particle-in-cell (2D PIC) simulations of double-pulse temporal shaping under a fixed total energy, jointly optimizing the prepulse energy fraction and the interpulse delay. The optimization process locates an asymmetric configuration with only about 7\% of the energy allocated to the leading pulse. Additionally, We also explain the physics by following the sequence from preplasma formation and laser absorption through hot-electron heating to rear sheath dynamics.

\section{Simulation method}\label{sec:methods}

\subsection{Particle-in-cell setup}\label{sec:pic}
The simulations are performed with the relativistic particle-in-cell code EPOCH in two dimensions~\cite{arber2015contemporary}. The target is a fully ionized CH foil of thickness $1\,\mu$m with electron density $n_e=184\,n_c$, where $n_c$ is the critical density for the drive wavelength. While two-dimensional (2D) modeling captures the dominant physics and scaling trends with high computational efficiency, it is well established that 2D simulations systematically overestimate the absolute proton cutoff energies compared to full three-dimensional (3D) simulations or experiments~\cite{babaei2017rise,sinigardi2018tnsa,dolier2022multi}. Consequently, our quantitative findings focus on the relative trends and the optimal parameter locations rather than the absolute energies. The drive is a linearly polarized Gaussian pulse with central wavelength $\lambda=0.8\,\mu$m, duration $\tau=25$~fs (full width at half maximum in intensity), and peak intensity $I_0=5.5\times10^{20}\,$W/cm$^2$, corresponding to a normalized amplitude $a_0\approx16$~\cite{esarey2009physics}. The grid spacing is $\Delta x=10$~nm along the laser axis, comparable to the collisionless skin depth at the foil density, and $\Delta y=24$~nm in the transverse direction. The computational domain and the number of macroparticles per cell follow standard practice for this class of problem.

\begin{figure}[t]
  \centering
  \includegraphics[width=1.0\columnwidth]{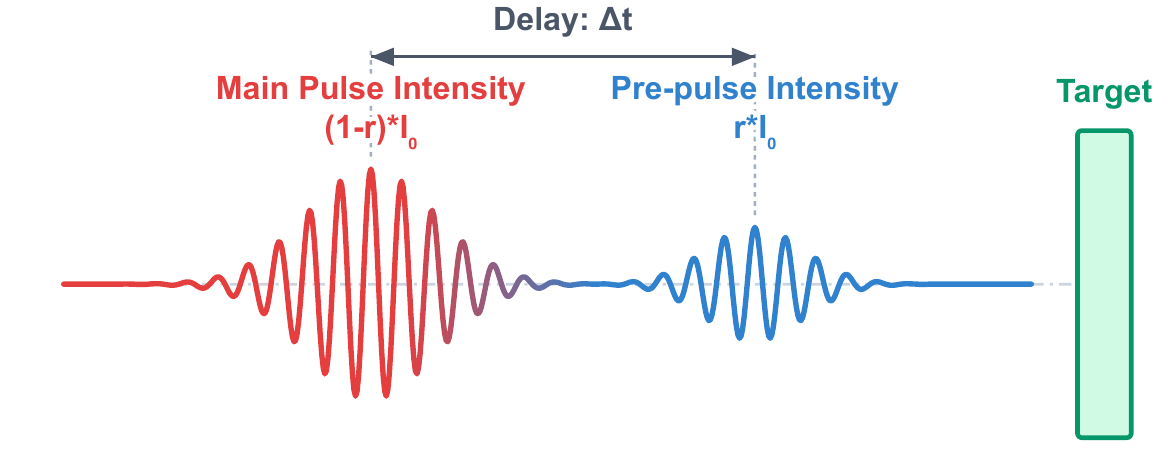}
  \caption{Schematic of the double-pulse scheme. The total drive energy is split
  between a leading pulse of peak intensity $r\,I_0$ and a main pulse of peak
  intensity $(1-r)\,I_0$, separated by a delay $\Delta t$ and incident on the
  foil.}
  \label{fig:schematic}
\end{figure}

\Cref{fig:schematic} illustrates the scheme. The single drive pulse is split into two collinear Gaussian pulses of equal duration $\tau=25$~fs. A fraction $r\in[0,1]$ of the total energy is placed in the leading pulse and the remaining fraction $1-r$ in the main pulse. Because the two pulses share the same spot size and duration, the peak intensities are
$r\,I_0$ and $(1-r)\,I_0$, so the total energy is conserved for any choice of
$r$. The second free parameter is the interpulse delay $\Delta t\in[50,350]$~fs,
measured between the pulse peaks. The single-pulse case is recovered at $r=0$ and serves as the baseline throughout.

The maximum proton energy $E_\mathrm{max}$ is extracted from the rear-directed
proton population ($p_x>0$, $x>1\,\mu$m). The baseline and the optimum are compared under an identical protocol, with the same total energy, the same drift
time for the accelerated protons, and a difference only in the prepulse fraction
($r=0$ versus $r\approx0.07$). Reported numbers and the mechanism figures use
single high-resolution reruns of the two endpoint cases. 

\subsection{Bayesian optimization method}\label{sec:bo}
The objective $E_\mathrm{max}(r,\Delta t)$ is treated as an expensive black-box
function. Each evaluation requires a full kinetic simulation, provides no
gradient information, and carries the statistical noise quoted above, so a
dense grid scan is impractical. Bayesian optimization is well suited to this
setting because it uses every completed simulation to build a model of the
objective and to decide where the next simulation is most informative,
typically reaching good solutions in far fewer evaluations than a systematic
scan~\cite{jones1998efficient,shahriari2016taking}.

The method has two ingredients, a probabilistic surrogate model and an
acquisition function. For the surrogate we use a Gaussian process, which places a
prior over functions defined by a mean and a covariance, or kernel. Conditioning
the prior on the evaluated points yields a posterior that gives, at any untried
point of the control space, both a predicted value and an uncertainty on that
prediction~\cite{rasmussen2006gaussian}. We use a Mat\'ern kernel with smoothness
parameter $\nu=5/2$, which corresponds to twice-differentiable sample functions
and is appropriate for a response that is smooth but not infinitely smooth. The
acquisition function then selects the next evaluation. We use an
upper-confidence-bound rule, which chooses the point that maximizes the posterior
mean plus a constant multiple of the posterior standard deviation, so that
regions of high predicted value and regions of high uncertainty are both
explored~\cite{srinivas2010gaussian}.

The campaign uses a fixed budget of 32 evaluations. The first 16 points are drawn
from a Sobol low-discrepancy sequence~\cite{sobol1967distribution}, which gives a space-filling coverage of the two-dimensional control space and initializes the surrogate before its predictions are relied upon. The remaining 16 points are selected adaptively by the acquisition function in two batches of eight. The control space $(r,\Delta t)$ is two-dimensional, which makes this budget sufficient to resolve the structure of the objective and to locate the optimum. The same approach has been used to optimize laser and plasma
parameters in related laser-plasma
studies~\cite{dolier2022multi,shalloo2020automation,jalas2021bayesian,loughran2023automated}.

During the optimization, each candidate is scored by a fixed-window objective in
which $E_\mathrm{max}$ is extracted 370~fs after the peak of the main pulse, the
same window for all 32 runs so that the objective ranks candidates consistently.
This window is chosen long enough that the high-field phase of the acceleration
has passed and the ordering of the cutoff energy across configurations has
stabilized, while keeping each evaluation inexpensive. At the two endpoints we
verified that extending the run to saturation raises the
absolute cutoff by at most about 2\%, within the simulation noise, and leaves the
ranking unchanged.

\section{Results and discussion}\label{sec:results}

\begin{figure}[t]
  \centering
  \includegraphics[width=1.0\columnwidth]{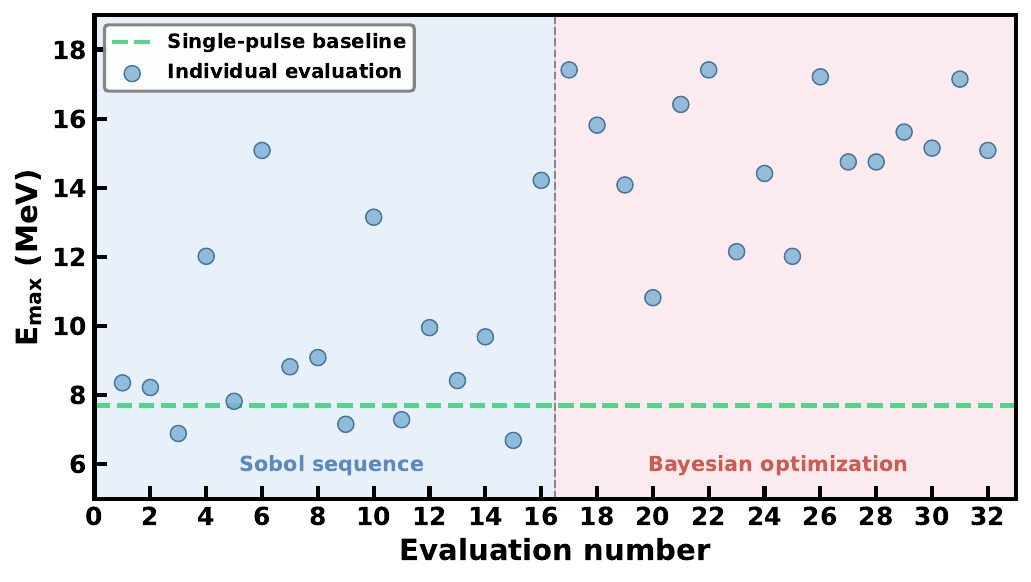}
  \caption{Optimization history showing the proton cutoff energy
  $E_\mathrm{max}$ versus evaluation number. Evaluations 1 to 16 (blue region)
  are Sobol-initialized, and evaluations 17 to 32 (red region) are adaptively
  selected by the Gaussian-process surrogate.}
  \label{fig:conv}
\end{figure}

\Cref{fig:conv} shows the optimization history, with the objective value of each
of the 32 evaluations plotted against its evaluation index. The 16
Sobol-initialized points (evaluations 1--16) sample the control space broadly and
span a wide range of objective values, reaching about 15~MeV at best. The 16
adaptive points chosen by Bayesian optimization (evaluations 17--32) concentrate
near the top of this range: the first adaptive point already enters the
high-performing region, and the later adaptive points remain within it, the best
objective value being about 17.4~MeV. 

\begin{figure}[t]
  \centering
  \includegraphics[width=1.05\columnwidth]{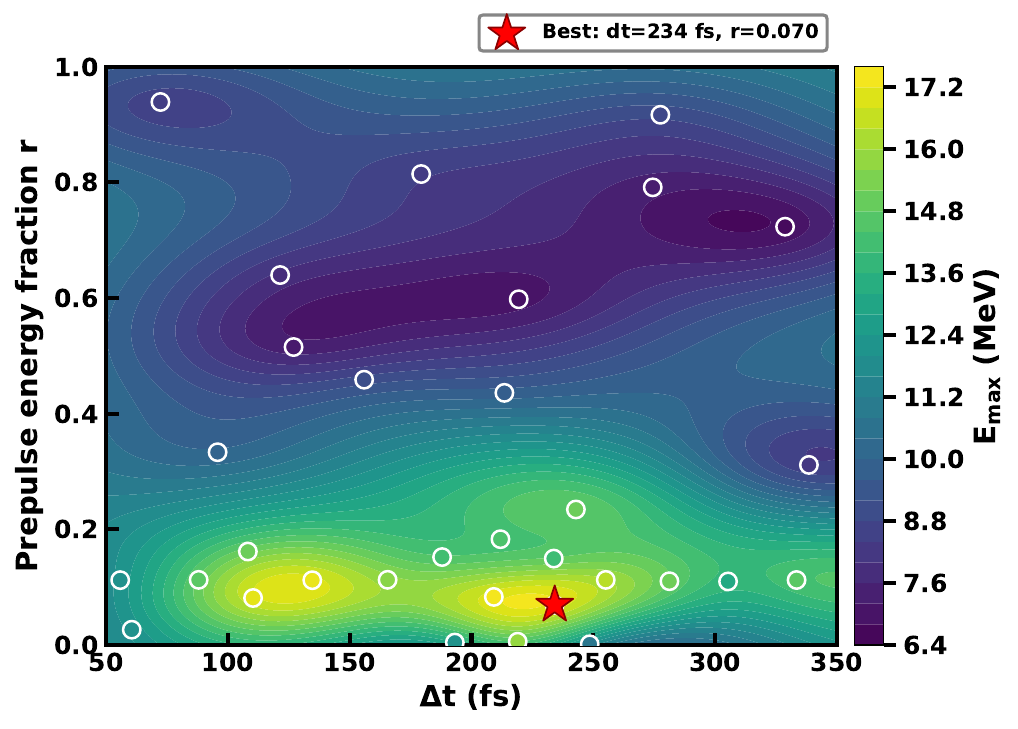}
  \caption{Gaussian-process posterior mean of $E_\mathrm{max}$ over the
  $(\Delta t,r)$ control space. Circles mark the 32 sampled points and the star
  marks the optimum ($r = 0.07$, $\Delta t = 234$~fs).}
  \label{fig:landscape}
\end{figure}

The Gaussian-process posterior mean over the control space is shown in
\cref{fig:landscape}, with the 32 sampled points and the optimum overlaid. This map shows the surrogate interpolation of the evaluations. The
high-$E_\mathrm{max}$ region is confined to small prepulse fractions,
$r = 0.05$ to $0.15$, and intermediate delays, $\Delta t = 100$ to
$250$~fs. Large prepulse fractions ($r>0.3$) give uniformly lower cutoff
energies, consistent with the trade-off imposed by the fixed energy budget,
because beyond a small fraction the energy removed from the main pulse outweighs
the benefit of a larger preplasma. 

\begin{figure}[t]
  \centering
  \includegraphics[width=1.0\columnwidth]{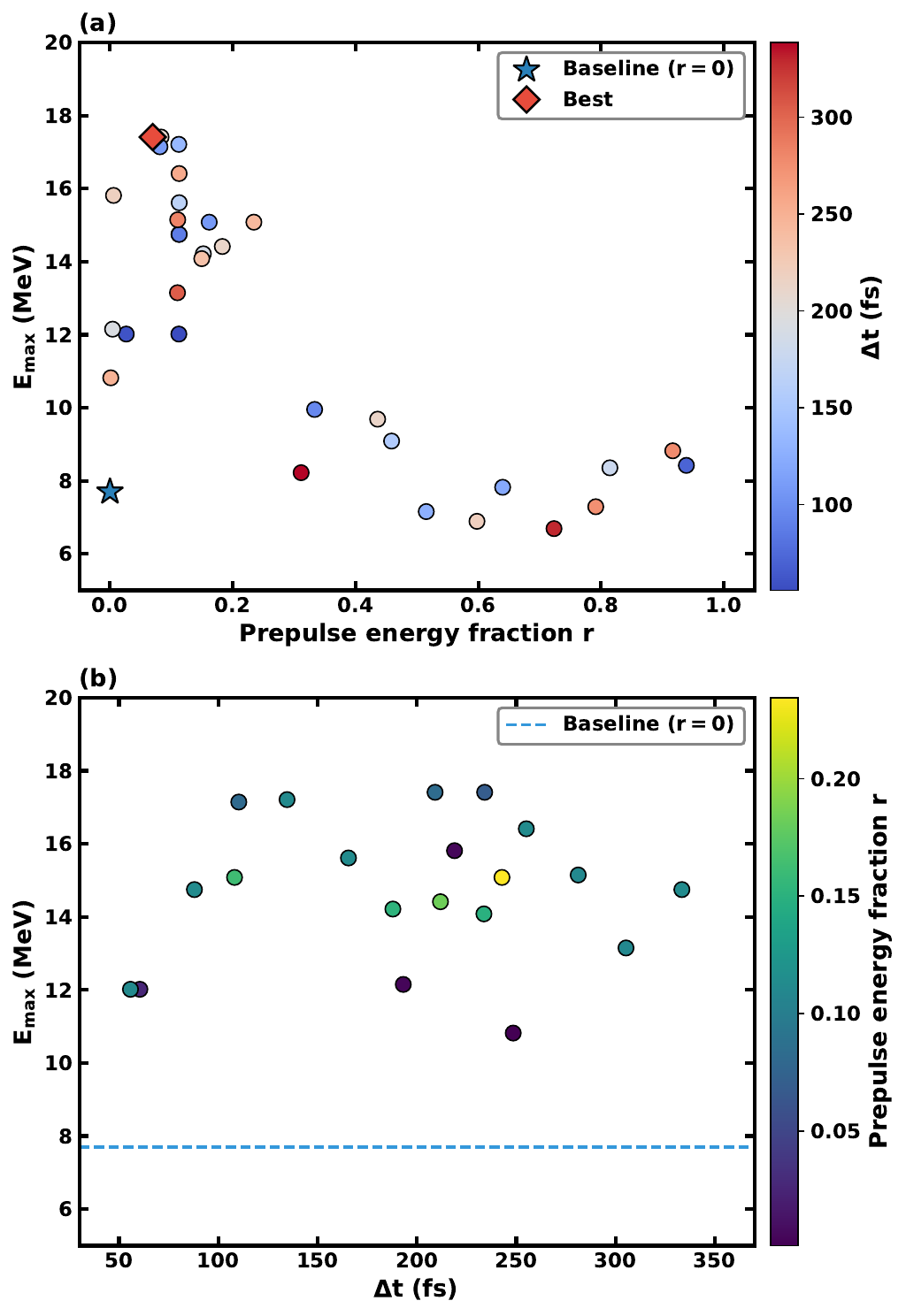}
  \caption{Parameter dependence of the cutoff energy. (a)~$E_\mathrm{max}$ versus
  prepulse fraction $r$ for all 32 evaluations, colored by delay $\Delta t$, with
  the baseline ($r=0$, 7.7~MeV) and the optimum marked. (b)~$E_\mathrm{max}$ versus delay $\Delta t$ for the
  near-optimal subset ($r<0.25$), colored by $r$, with the dashed line marking
  the baseline.}
  \label{fig:slices}
\end{figure}

The dependence on each parameter is shown in \cref{fig:slices}. Panel~(a) plots
$E_\mathrm{max}$ against $r$ for all 32 evaluations, colored by $\Delta t$. The
cutoff energy is highest in the sweet spot $r\in[0.05,0.15]$ and decreases
monotonically as $r$ increases further. Across the campaign the cutoff energy and
$r$ are strongly anticorrelated, with a correlation coefficient of $-0.792$, and
every double-pulse point lies above the single-pulse baseline. Panel~(b) plots
$E_\mathrm{max}$ against $\Delta t$ for the near-optimal subset ($r<0.25$),
colored by $r$. Here the cutoff energy is nearly flat over a broad plateau,
$\Delta t\in[100,250]$~fs, with all points above 12~MeV. The optimum sits on a plateau in delay, which is favorable for an experiment because it relaxes the timing tolerance.

\begin{figure}[t]
  \centering
  \includegraphics[width=1.0\columnwidth]{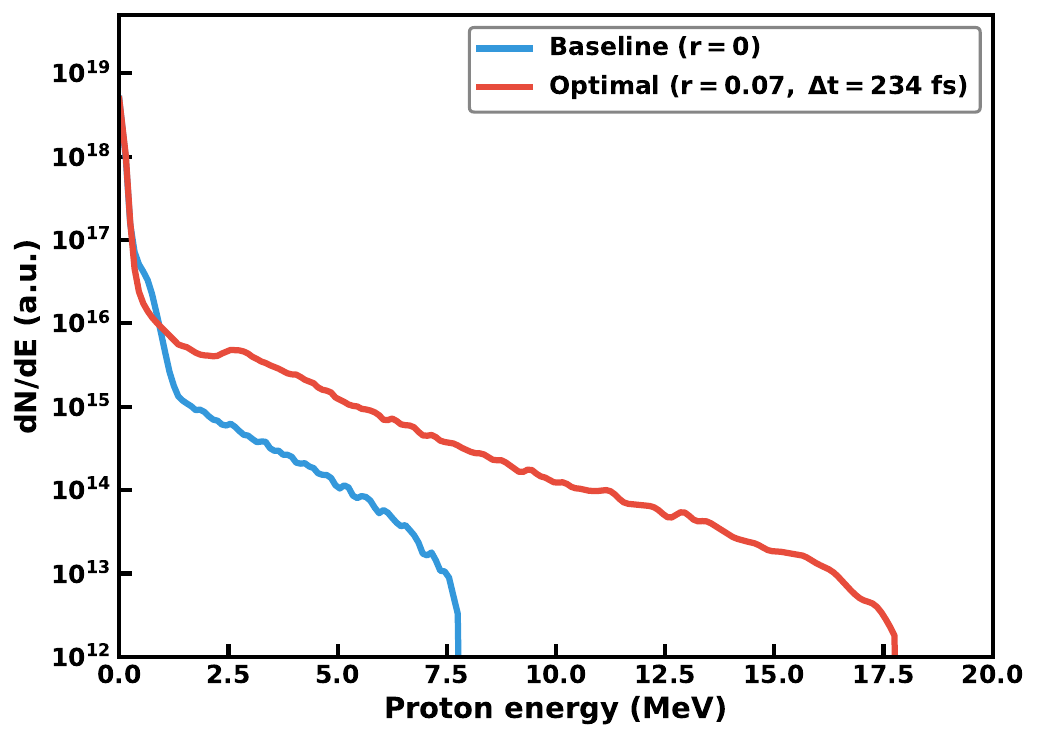}
  \caption{Rear-directed proton energy spectra $\mathrm{d}N/\mathrm{d}E$ for the
  single-pulse baseline and the optimized double pulse at $t=1000$~fs.}
  \label{fig:spectrum}
\end{figure}

The proton energy spectra at the baseline and the optimum are compared in
\cref{fig:spectrum} using the endpoint reruns. The single-pulse spectrum has a
cutoff of 7.7~MeV. The optimized double pulse extends the cutoff to 17.7~MeV, a
gain of about 130\%, and also raises the yield at
intermediate energies. This near-doubling of the cutoff energy is achieved under a fixed energy budget, it may originate in how the asymmetric double pulse couples to the target rather than in any additional drive energy. To identify its origin, we follow the mechanism in the order in which it acts, beginning with the front-surface density that the main pulse encounters and proceeding through the laser absorption, the electron heating, and the rear sheath field.

\begin{figure}[t]
  \centering
  \includegraphics[width=1.0\columnwidth]{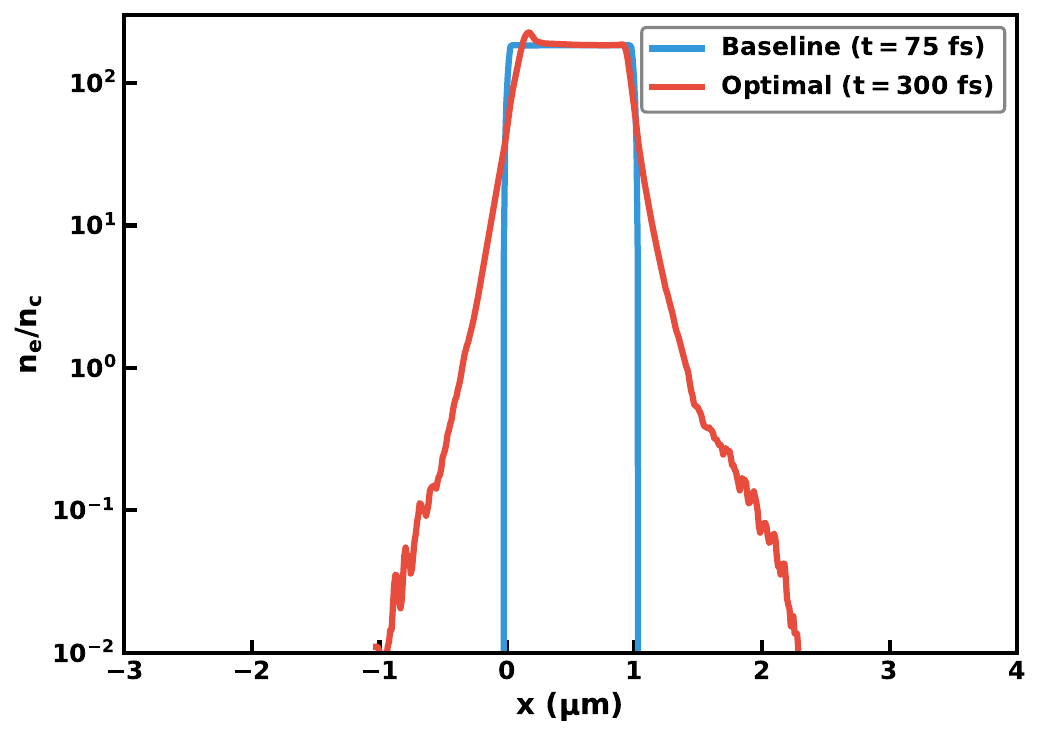}
  \caption{Electron density profiles $n_e/n_c$ along the laser axis for the
  baseline ($t=75$~fs, before the single pulse) and the optimum
  ($t=300$~fs, before the main pulse).}
  \label{fig:density}
\end{figure}

\Cref{fig:density} compares the on-axis electron density profile that the main
pulse encounters in the two cases. Each profile is taken at the last instant
before the main pulse reaches the target, so that the comparison is made at the
same physical stage of the interaction. For the single-pulse baseline this instant
is $t=75$~fs, just before the only pulse arrives; for the optimum it is
$t=300$~fs, after the leading pulse and the delay $\Delta t$ but before the main
pulse. Because the baseline target has not yet been irradiated at $t=75$~fs, its
profile is simply the initial, unperturbed top-hat distribution, with sharp
boundaries at $x=0$ and $x=1\,\mu$m and a flat interior at $n_e=184\,n_c$. In the
optimum, by contrast, the leading pulse has already heated and expanded the front
surface during the interval $\Delta t$, so that the main pulse instead meets an
approximately exponential preplasma ramp extending about $2\,\mu$m in front of the
original surface. The weak leading pulse ablates only a thin surface layer, whose expansion into vacuum forms the exponential ramp, so the dense core retains its initial height while the low-density wings of the ramp contain only a small fraction of the target electrons; the modest rise of the peak above $184\,n_c$ reflects compression of the surface by the ablation pressure. The rear surface remains steep at this time, so the sharp density gradient required for sheath acceleration is preserved.

\begin{figure*}[t]
  \centering
  \includegraphics[width=1.02\textwidth]{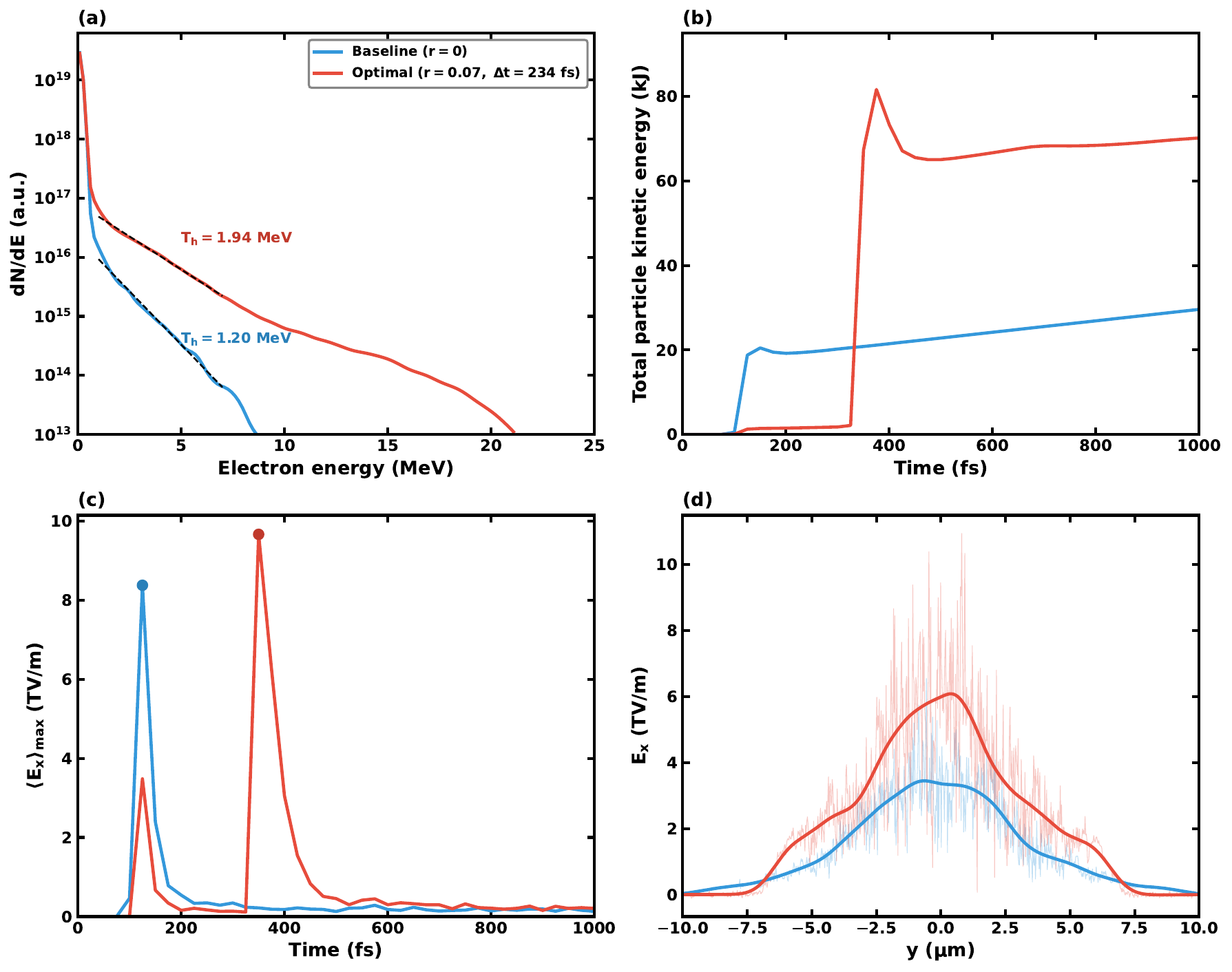}
  \caption{Mechanism diagnostics for the baseline and the optimum. 
  (a)~Hot-electron spectra sampled after each main pulse, with single-temperature fits over the same $[1,7]$~MeV window giving $T_h=1.20$~MeV (baseline) and $1.94$~MeV (optimum). 
  (b)~Temporal evolution of the total particle kinetic energy in the simulation box.
  (c)~Rear sheath field versus time, from the transversely averaged $E_x$ behind the foil. 
  (d)~Transverse profile of the rear field at the time of peak field.}
  \label{fig:mechanism}
\end{figure*}

The relaxed front gradient increases the coupling of the main pulse to the plasma, generating a hotter electron population. 
\Cref{fig:mechanism}(a) shows the hot-electron spectra sampled just after each main pulse has finished interacting with the target. A single-temperature fit over the same $[1,7]$~MeV interval gives a bulk hot-electron temperature of $T_h = 1.20$~MeV in the baseline and $1.94$~MeV at the optimum, representing an increase of about 61\%. 

This hotter electron population is driven by the enhanced energy transfer from the laser to the plasma. To quantify this energy coupling, the laser energy absorption fraction $\eta_{\mathrm{abs}}$ is calculated by integrating the Poynting vector flux over the boundaries of the simulation domain $\partial\Omega$ over the entire simulation duration $t_{\mathrm{end}} = 1000$~fs:
\begin{equation}
\eta_{\mathrm{abs}} = \frac{E_{\mathrm{inj}} - E_{\mathrm{esc}}}{E_{\mathrm{inj}}},
\label{eq:absorption}
\end{equation}
where $E_{\mathrm{inj}} = \int_{0}^{t_{\mathrm{end}}} \oint_{\partial \Omega} (\mathbf{E} \times \mathbf{H})_{\mathrm{inj}} \cdot \mathrm{d}\mathbf{S} \mathrm{d}t$ is the total laser energy injected into the box, and $E_{\mathrm{esc}} = \int_{0}^{t_{\mathrm{end}}} \oint_{\partial \Omega} (\mathbf{E} \times \mathbf{H})_{\mathrm{esc}} \cdot \mathrm{d}\mathbf{S} \mathrm{d}t$ is the total electromagnetic energy that escapes through the boundaries (representing reflection, transmission, and scattering). Here, $\mathrm{d}\mathbf{S} = \hat{\mathbf{n}}\,\mathrm{d}S$ is the outward-pointing vector surface element of the boundary $\partial\Omega$, where $\mathrm{d}S$ is the local area element and $\hat{\mathbf{n}}$ the local outward unit normal.
For both the baseline and optimized cases, the total injected energy is identical at $E_{\mathrm{inj}} \approx 389.1$~kJ. In the baseline case, the total escaped electromagnetic energy is $E_{\mathrm{esc}} \approx 370.3$~kJ, which corresponds to an absorption fraction of $\eta_{\mathrm{abs}} \approx 4.84\%$ (equivalent to an absorbed energy of $E_{\mathrm{abs}} \approx 18.8$~kJ). In the optimized case, the preplasma ramp suppresses reflection, reducing the escaped energy to $E_{\mathrm{esc}} \approx 310.9$~kJ. This yields a substantial increase in the absorption fraction to $\eta_{\mathrm{abs}} \approx 20.09\%$ (absorbed energy $E_{\mathrm{abs}} \approx 78.2$~kJ), representing a 4.15-fold enhancement in energy coupling.

This energy transfer is visible in the temporal evolution of the total particle kinetic energy inside the simulation box, $E_{\mathrm{particle}}(t)$, shown in \cref{fig:mechanism}(b). For the baseline case, heating occurs in a single stage starting at $t \approx 100$~fs as the pulse hits the sharp target front surface. The particle kinetic energy inside the box initially rises to $\sim 19.6$~kJ, which is bounded by the transient absorbed laser energy of $\approx 21.9$~kJ at that moment ($t \approx 200$~fs, before the reflected waves fully escape the simulation boundaries), and subsequently exhibits a slow, linear increase to $29.6$~kJ at $t = 1000$~fs due to cumulative numerical grid heating. In contrast, the optimized case displays a clear two-stage heating process: a minor prepulse heating stage starting at $t \approx 100$~fs, which transfers about $1.4$~kJ to the plasma to form the preplasma ramp, followed by a major heating stage at $t \approx 320$~fs upon the arrival of the main pulse. The total particle energy peaks at approximately $81.7$~kJ at $t \approx 380$~fs, well below the transient absorbed laser energy of $\approx 97.2$~kJ at the same instant, thereby strictly satisfying energy conservation. It then decreases to $70.2$~kJ by $t=1000$~fs as high-energy electrons escape from the simulation box boundaries, carrying away their kinetic energy.

The more energetic electron population produces a stronger and longer-lived rear sheath. \Cref{fig:mechanism}(c) shows the rear sheath field, obtained by transversely averaging $E_x$ behind the foil to remove filamentary noise and then taking its peak over the transverse coordinate. The peak field rises only modestly, from 8.4 to 9.7~TV/m, an increase of 15\%, which alone does not account for the factor of about 2.3 in proton energy. The relevant quantities are instead the time-integrated field and the transverse extent of the accelerating region. The time integral of the sheath field (sheath impulse) increases from 474 to 825~TV/m$\cdot$fs, a factor of about 1.7. 

Finally, the transverse extent of the accelerating region is shown in \cref{fig:mechanism}(d), which compares the transverse profiles of the rear field $E_x(y)$ at the time of peak field, using the same transverse averaging as panel (c). The optimized profile is broader (covering approximately $\pm 6\,\mu\mathrm{m}$ compared to $\pm 4\,\mu\mathrm{m}$ in the baseline), indicating that the sheath field is more persistent and covers a larger area rather than being much higher at its peak. The increase in proton cutoff energy follows from this more sustained and transversely wider sheath acceleration. In the TNSA mechanism, the final cutoff energy of accelerated protons is determined by the integration of the accelerating field along the ion trajectories, $E_{\mathrm{max}} \approx \int e E_x(x(t), y(t), t) \, \mathrm{d}t$. Although the peak sheath field strength is only enhanced by 15\%, the 1.7-fold increase in the sheath impulse combined with a broader transverse acceleration area provides a significantly more durable and stable accelerating environment. This prevents the rapid decay of the accelerating field and allows the protons to gain energy over a longer distance and duration, leading to the substantial 130\% increase in the cutoff energy observed in the optimized double-pulse configuration.

\section{Conclusion}\label{sec:conclusion}
We have optimized double-pulse temporal shaping for TNSA proton acceleration under
a fixed total laser energy, treating the prepulse energy fraction and the
interpulse delay as coupled control parameters and using Bayesian optimization to
search the two-dimensional space efficiently. A campaign of 32 simulations locates
an asymmetric optimum at $r\approx0.07$ and $\Delta t\approx234$~fs, at which the
proton cutoff energy increases from 7.7~MeV to 17.7~MeV, a gain of about 130\%.
The gain follows a quantifiable chain in which the leading pulse forms a front-surface preplasma, the main pulse couples into it with absorption rising from 4.84\% to 20.09\%, the hot-electron temperature increases from 1.20 to 1.94~MeV, and the rear sheath becomes stronger, more persistent, and transversely broader, with its time integral increasing by a factor of about 1.7. The optimum lies on a broad delay plateau, which relaxes the experimental timing tolerance. The absolute energies are subject to the overestimate inherent in two-dimensional modelling, and three-dimensional simulations together with an experimental test of the asymmetric optimum are natural next steps.

\begin{acknowledgments}
This work is supported by the National Natural Science Foundation of China (NSFC) under Grants No.12375240, No.12535015 and the Program of China Scholarship Council (Grant CSC202506040219).
\end{acknowledgments}

\section*{Conflict of Interest}
The authors declare no conflicts of interest.

\section*{Data availability}
The data that support the findings of this study are available from the corresponding author upon reasonable request.

\section*{Author Contributions}
\textbf{Cheng-Qi Zhang:} Conceptualization (lead); Methodology (lead); Software (lead); Investigation (lead); Formal analysis (lead); Data curation (lead); Writing original draft (lead); Writing review \& editing (lead). \\
\textbf{Yang He:} Conceptualization (supporting); Methodology (supporting); Writing review \& editing (supporting). \\
\textbf{Mamat Ali Bake:} Formal analysis (supporting); Writing review \& editing (supporting). \\
\textbf{Bai-Song Xie:} Supervision (lead); Funding acquisition (lead); 
Writing -- review \& editing (supporting).

\bibliographystyle{unsrt}
\bibliography{ref}

\end{document}